# The Roadmap of Inorganic Computational Materials Databases: Capabilities, Credibility, Coverage, and the Open Frontier

Miao Liu[1,2]*, Jianghao Jin[1,2], Tenglong Lu[1,2], Jianguo Si[2], Yin Shi[1], Sheng Meng[2]*, Weihua Wang[2]*

Institute of Physics, Chinese Academy of Sciences, Beijing 100190, China;

Dongguan Institute of Materials Science and Technology, Dongguan, Guangdong 523808,China

* E-mail: mliu@iphy.ac.cn, smeng@iphy.ac.cn, whw@dimst.ac.cn

**Abstract**

Computational materials databases have become central infrastructure for data-driven discovery of inorganic materials, yet their growth remains strikingly uneven across property families.This perspective synthesizes a systematic survey of mainstream density functional theory (DFT) software, the computational cost and credibility of nineteen material-property families, and the coverage of existing computational databases, into a coherent picture of where the field stands and where it should go.We show that the ecosystem of first-principles codes is methodologically mature: for nearly every property of technological interest, at least one production-grade code can compute it.The binding constraint is no longer methodological capability but the economics of trust — which properties can be computed cheaply enough, and accurately enough, to be harvested at database scale. Mapping database coverage onto a Gartner-style readiness cycle reveals a sharp divide: ground-state structure, energetics, elasticity, and topology have reached routine production, while nine property families — including NMR/EPR parameters, core-level spectra, electron–phonon properties, thermal conductivity, and quantum transport — remain without any systematic computational database.We argue that these blank zones define the scientific opportunity of the next decade, and we propose a three-horizon roadmap: consolidating coverage and interoperability in the near term, industrializing mid-cost properties through surrogate-accelerated workflows in the medium term, and conquering the high-cost frontier through machine-learned interatomic potentials, autonomous computing infrastructure, and community governance in the long term.

## 1. Introduction

The discovery of inorganic materials has been transformed over the past fifteen years by a simple idea: instead of computing properties one material at a time, compute them for entire chemical spaces and publish the results as searchable databases [1–5].

The Materials Project (MP), the Open Quantum Materials Database (OQMD), AFLOW, JARVIS, and Atomly now collectively hold millions of computed compounds, underpinning modern machine-learning interatomic potentials [6–8], generative models for crystal structures [6], and autonomous "self-driving" laboratories.

What began as a materials-genome demonstrator has become the reference infrastructure of an entire discipline, consulted daily by experimentalists who have never run a first-principles calculation themselves.

This article focuses deliberately on inorganic computational databases of properties derived from quantum-mechanical — overwhelmingly density functional theory (DFT)[9,10]— calculations.

Every entry in such a database is the deterministic output of a well-defined computation, which means that coverage gaps are never accidents of history — they are decisions, forced or otherwise, about what is worth computing, what can be computed affordably, and what can be trusted.

And those decisions have produced a curious asymmetry: formation energies and lattice parameters are abundant, while for a converged phonon spectrum, a hybrid-functional band gap, or an NMR chemical shift, coverage thins out or vanishes entirely.

The asymmetry is not accidental, nor is it primarily a failure of software. It reflects the interplay of three forces that this article dissects in turn: what the available codes can compute (capability), what it costs and how much one can trust the result (cost and credibility), and what the community has actually chosen to compute and publish (coverage).

Understanding this interplay matters now more than ever. The community is actively debating what the next generation of materials databases should contain, and several large-scale initiatives are being planned around automated computation and AI-assisted characterization; decisions about which property families to industrialize next will shape a decade of infrastructure investment. This article aims to inform those decisions.

We deliberately keep the discussion accessible, so that a materials scientist who has never run a DFT calculation and a database architect planning the next platform can both follow the argument.

The analysis rests on three figures and one table developed from a systematic survey of code documentation, published high-throughput studies, and the public pages of major databases (status 2025–2026). Together they support a single claim: the frontier of inorganic computational databases has moved from *can we compute* to *can we afford to compute at scale, and can we trust what we publish* — and the roadmap of the field should be drawn accordingly.

## 2. What can DFT codes calculate?

Density functional theory is, by most measures, the most successful method in computational materials science. Introduced by Hohenberg and Kohn in 1964 [9] and cast into practical form by Kohn and Sham a year later [10], it recasts the intractable many-electron problem as a self-consistent single-particle problem for the electron density alone — a reformulation whose development earned the 1998 Nobel Prize in Chemistry.

Modern semilocal functionals such as PBE [11] predict lattice parameters to within 1–2% and formation energies to within roughly 0.1 eV per atom; hybrid functionals [12] and many-body corrections extend this accuracy to band gaps and excited states.

The result is a method that is simultaneously first-principles (no system-specific fitting), chemically general, and affordable on routine computing clusters.

Its adoption is correspondingly vast: flagship codes count their users in the tens of thousands and their citations in the hundreds of thousands, and essentially every quantitative statement in modern materials physics — a lattice constant, a phonon spectrum, a hyperfine field — can be, and usually has been, computed by DFT.

The bibliometric record makes the point unusually concrete.

The two most-cited papers in the century-long history of the Physical Review journals are the founding papers of the theory itself [9,10,13]; the single most-cited Physical Review Letter is the PBE functional [11,14]; as of 2021, all ten of the most-cited articles ever published by the American Physical Society concern DFT or its practical apparatus — pseudopotentials, PAW, gradient corrections — [14]; and 12 of the 100 most-cited papers of all time, across all of science, are DFT methods papers [15].

No other method in the physical sciences approaches this footprint.

What, concretely, can these codes calculate? The short answer is: nearly every property of a crystalline solid that an experimentalist can measure.

From a single input structure, a modern code returns the relaxed geometry, total energy, and forces — and from these follow formation energies, phase stability, elastic constants, and the equation of state. Layered on the same ground state are the excited-state methods: band structures and band gaps, dielectric and piezoelectric tensors, and optical spectra.

Density-functional perturbation theory (DFPT) [16] adds phonon dispersions and lattice thermal properties; electron–phonon methods [17] add carrier mobilities and superconducting critical temperatures; all-electron and GIPAW routes add NMR chemical shifts and electric field gradients; and specialized modules extend the reach to quantum transport and ab initio molecular dynamics.

For a database project, therefore, the question is no longer *whether* a property can be computed, but *which code should compute it* — and that choice is far from trivial.

The mainstream ecosystem consists of plane-wave pseudopotential codes (VASP [18], Quantum ESPRESSO [19,20], CASTEP [21], ABINIT [22]), all-electron codes (WIEN2k [23]), local-orbital codes (SIESTA [24]), and Gaussian/plane-wave hybrids (CP2K [25]), each with distinct methodological strengths.

Figure 1 condenses the selection logic into a decision tree. Reading the tree from the root, the first branch separates ground-state and thermodynamic properties from electronic-structure properties such as band gaps and optical spectra, from dynamical properties such as phonons, and from specialized spectroscopies. The leaves then encode hard-won community experience.

VASP, with its PAW formalism and robust DFPT implementation [16], is the default workhorse for most bulk properties. Quantum ESPRESSO offers the deepest open-source DFPT stack and, through the EPW code [17], the only practical route to electron–phonon couplings, carrier mobilities, and superconducting critical temperatures.

WIEN2k, as an all-electron code, is the reference for properties that depend on the core region — NMR chemical shifts and electric field gradients — and its TB-mBJ potential delivers band gaps of near-hybrid quality at a fraction of the hybrid cost. CASTEP is the natural choice for NMR in molecular and framework solids; SIESTA, through TranSIESTA, dominates quantum transport calculations; CP2K excels at ab initio molecular dynamics (AIMD) on large systems.

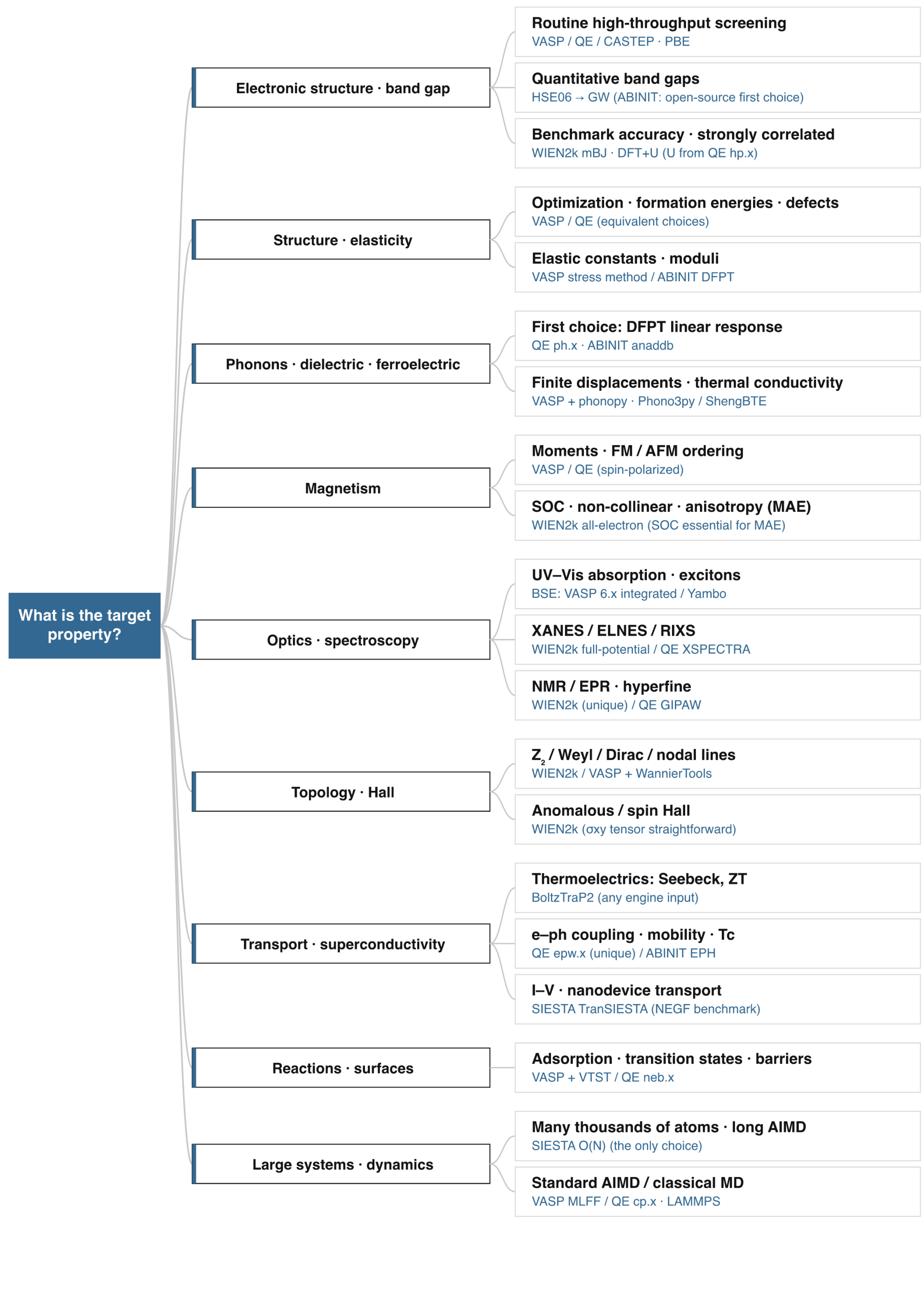

Figure 1 | Decision tree for DFT code selection. Starting from the research question at the root, nine branches lead to twenty recommended code–method combinations, annotated with typical computational cost and accuracy caveats.

Table 1 | Capability matrix of six mainstream DFT codes across nineteen property families. ★ best-in-class or unique capability; ● mature, benchmark-validated implementation; ○ possible but

limited, or reliant on external tools; — not supported. Annotations name the relevant module or method; the final row records the license model.

| Property family | VASP | QE | CASTEP | WIEN2k | ABINIT | SIESTA |
|---|---|---|---|---|---|---|
| Structure · energy · forces | ● | ● | ● | ● | ● | ● |
| Band structure · gap (high accuracy) | ● HSE/GW | ● HSE/Yambo | ● HSE/mBJ | ★ mBJ | ★ best open GW | ○ |
| Phonons · dielectric · ferroelectric | ○ +Phonopy | ★ ph.x DFPT | ● finite disp. | ○ interfaces | ★ anaddb | ○ |
| Lattice thermal conductivity | ○ Phono3py | ○ Phono3py | ○ Phono3py | ○ | ○ Phono3py | — |
| Elastic constants | ● | ● | ● | ● | ● | ○ |
| Magnetism · SOC · MAE | ● | ● | ● | ★ all-electron | ● | ● |
| Anomalous / spin Hall | ○ | ○ | — | ★ σxy tensor | ○ | — |
| Topology (Z2 · Weyl · nodal lines) | ● WannierTools | ○ Wannier90 | ○ | ★ Berry curvature | ○ Wannier90 | — |
| Optical dielectric ε(ω) | ● LOPTICS | ● epsilon.x | ● incl. phonons | ★ all-electron | ● | ○ |
| BSE excitons | ★ integrated | ● Yambo | ○ | ● | ● built-in | ○ low accuracy |
| XANES / ELNES / RIXS | ● | ● XSPECTRA | ○ | ★ essential | ○ | — |
| NMR / EPR · hyperfine | ○ | ● GIPAW | ● | ★ unique | ○ | — |
| Thermoelectric Seebeck · ZT | ○ BoltzTraP2 | ○ BoltzTraP2 | ○ | ● native | ○ | ○ |
| e–ph coupling · mobility · Tc | ○ | ★ epw.x unique | — | — | ● EPH | — |
| Quantum transport (NEGF · I–V) | — | ● PWcond | — | — | — | ★ TranSIESTA |
| Reaction paths · NEB | ● +VTST | ● neb.x | ● | ○ | ● | ● |
| High-pressure transitions | ○ | ○ | ○ | ● | ★ native | — |
| Strongly correlated (+U · DMFT) | ● +U | ★ hp.x auto-U | ● +U | ● +U/DMFT | ● DMFT | ○ +U |
| Large-scale AIMD · long times | ● MLFF | ○ | ○ | — | ○ | ★ O(N) unique |
| **License** | **Commercial** | **Open source** | **Commercial (MS)** | **Commercial (€400)** | **Open source** | **Open source** |

Table 1 generalizes the tree into a capability matrix: six codes against nineteen property families, graded from native, benchmark-validated support down to "not available." The breadth of the computable is striking — yet the matrix also carries a sobering message.

Every cell condenses a chain of methodological decisions — which functional rung, which pseudopotential or all-electron treatment, which convergence protocol, which module of which code — and each of those decisions conditions the accuracy, the cost, and ultimately the credibility of the number that ends up in a database.

Building a high-throughput database therefore presupposes deep, practitioner-level understanding of DFT methodology and of the software that implements it; without that expertise, mass computation yields reproducible-looking numbers rather than trustworthy data.

The same expertise, moreover, must be multiplied rather than concentrated: no single code spans the matrix, so a comprehensive effort has to orchestrate several of them, with direct consequences for workflow engineering, data interoperability, and quality control.

The practical conclusion of this section is optimistic: the methodological substrate for a comprehensive inorganic computational database already exists. Whatever the next generation of databases chooses to compute, the tools to compute it are — with few exceptions — ready.

One caveat deserves emphasis, because it shapes everything that follows: the methodological layer beneath the matrix.

Plane-wave pseudopotential codes treat core electrons implicitly, which is why properties that probe the core region — hyperfine fields, chemical shifts, core-level spectra — systematically favor the all-electron or GIPAW routes. A database architect choosing a code for a property family is therefore also choosing an accuracy ceiling, and that ceiling must be documented as part of the data.

The real constraints appear when we ask what it costs to run these tools at scale, and how much confidence the results deserve.

## 3. The economics of trust: cost versus credibility

Not all computed properties are born equal. A formation energy from a well-converged PBE calculation costs tens of core-hours and typically agrees with experiment to within 50–100 meV/atom; a superconducting critical temperature from an EPW workflow can cost $10^5$–$10^6$ core-hours and still carries methodological uncertainties that are difficult to bound.

Figure 2 arranges the nineteen property families on this two-dimensional landscape — computational credibility along the horizontal axis, computational cost along the vertical, with cheaper calculations sitting higher — while the marker fill records what existing databases have already harvested (solid for large-scale coverage, half-filled for partial, hollow for none).

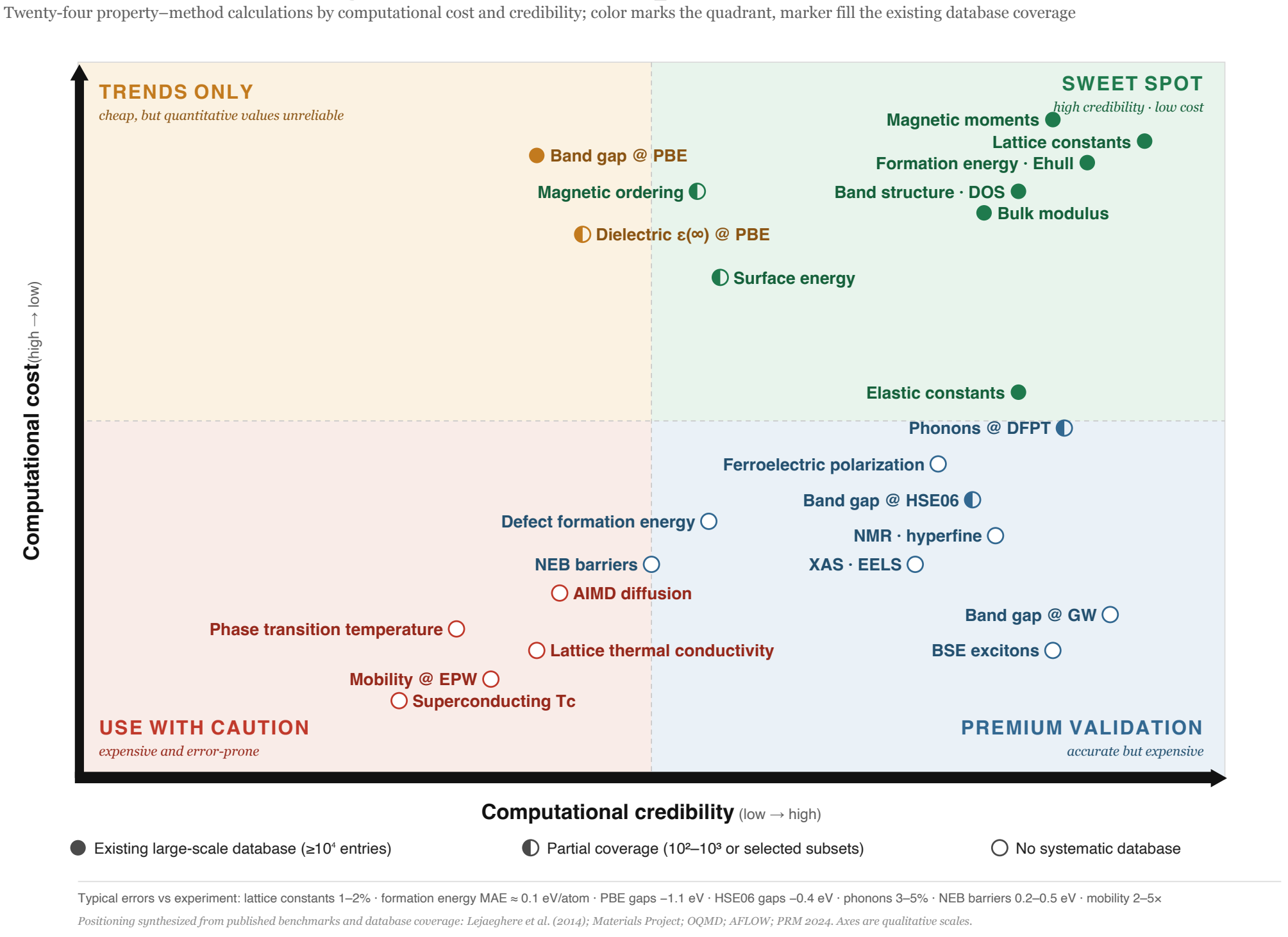

Figure 2 | Cost–credibility quadrant for nineteen computable property families. The upper-right "sweet spot" combines high credibility with low cost; marker fill encodes the coverage of existing computational databases (solid, large-scale; half-filled, partial; hollow, none).

The figure invites a calm reading. The upper-right corner — the sweet spot where credibility is high and cost is low — is occupied by exactly the simple, robust, easy-to-use methods that the community has already taken up broadly: lattice parameters, formation energies, elastic constants, magnetic moments. The first generation of databases, in other words, harvested what the quadrant says was easiest to harvest.

Read further, the figure offers a quiet form of guidance for the road ahead: it shows at a glance which property families are both cheap and trustworthy — the natural candidates for the next round of database building — and which ones will ask for more, whether in computing budget, methodological care, or patience. It is less a verdict on the field than a compass for planning one: property by property, it locates where data can still be grown with the least friction.

Some of the spread, it should be said, belongs to the method itself rather than to any implementation. The canonical example is the band gap: semilocal functionals such as PBE [11] systematically underestimate experimental gaps by 30–50%, and no amount of additional compute removes that bias — it is a functional error, not a convergence error.

Higher rungs of the ladder can close it: hybrid functionals such as HSE06 [12] and GW bring gaps to within a few tenths of an electronvolt of experiment, at one to two orders of magnitude greater cost. DFT has always been an art of balance.

Every property in the figure embodies a trade-off between accuracy and efficiency, and its position simply records where that balance has been struck so far — cheap, approximate protocols are screened at scale, expensive, accurate ones are reserved for curated subsets, and the craft of database design lies in matching the rung of the ladder to the purpose the data will serve.

This balance is not an abstraction; it is the working arithmetic of real discovery campaigns. Screening a battery cathode or a solid-state electrolyte, for instance, means ranking candidates by thermodynamic stability, electrochemical window, and — the expensive part — ionic mobility.

The FastTrack framework of Kang et al. [26] attacks precisely this bottleneck: by combining universal machine-learned interatomic potentials with three-dimensional potential-energy-surface sampling, it evaluates migration barriers within tens of millielectronvolts of DFT nudged-elastic-band results while running roughly two orders of magnitude faster, turning ion-transport screening from a per-material project into a database-scale operation.

Building on the Atomly database  of first-principles properties, Lu et al. [27]screened 54,005 compounds for thermodynamic stability, electrochemical and chemical compatibility, electronic insulation, and ionic conductance, identifying 41 coating materials that stabilize the $LiFePO_4$–sulfide-electrolyte interface in all-solid-state batteries.

The same logic reaches beyond energy materials: Yu et al. [28]screened approximately 182,000 structures for $MgB_2$-like superconductors, shortlisted five candidate families, and synthesized one of them — $BaGa_2$, experimentally confirmed as a BCS superconductor with Tc = 0.36 K, in good agreement with the theoretical prediction.

Each campaign is, at bottom, a database query: properties cheap enough and trustworthy enough to compute at scale, composed into a filter that only a handful of materials survive.

The rest of the landscape follows the same logic. A band of trustworthy but expensive properties — hybrid and GW band gaps, full phonon dispersions, dielectric and piezoelectric tensors — is computed today only for curated subsets, awaiting the cost reductions that surrogate models and automated workflows are beginning to deliver.

A further band, delicate as well as costly — electron–phonon properties, NMR parameters in complex solids, quantum transport — has not yet found its balance: its results depend on methodological choices in ways that make large-scale publication premature without careful validation. Neither region is closed; both are simply waiting for the economics to turn in their favor, a theme the next two sections develop.

## 4. What has been harvested: coverage and readiness

Figures 1–2 and Table 1 concern what *could* be computed. Figure 3 turns to what *has* been computed and published, arranging the nineteen property families along a hype-cycle-style "data readiness curve" whose horizontal axis measures the maturity of the corresponding data ecosystem.

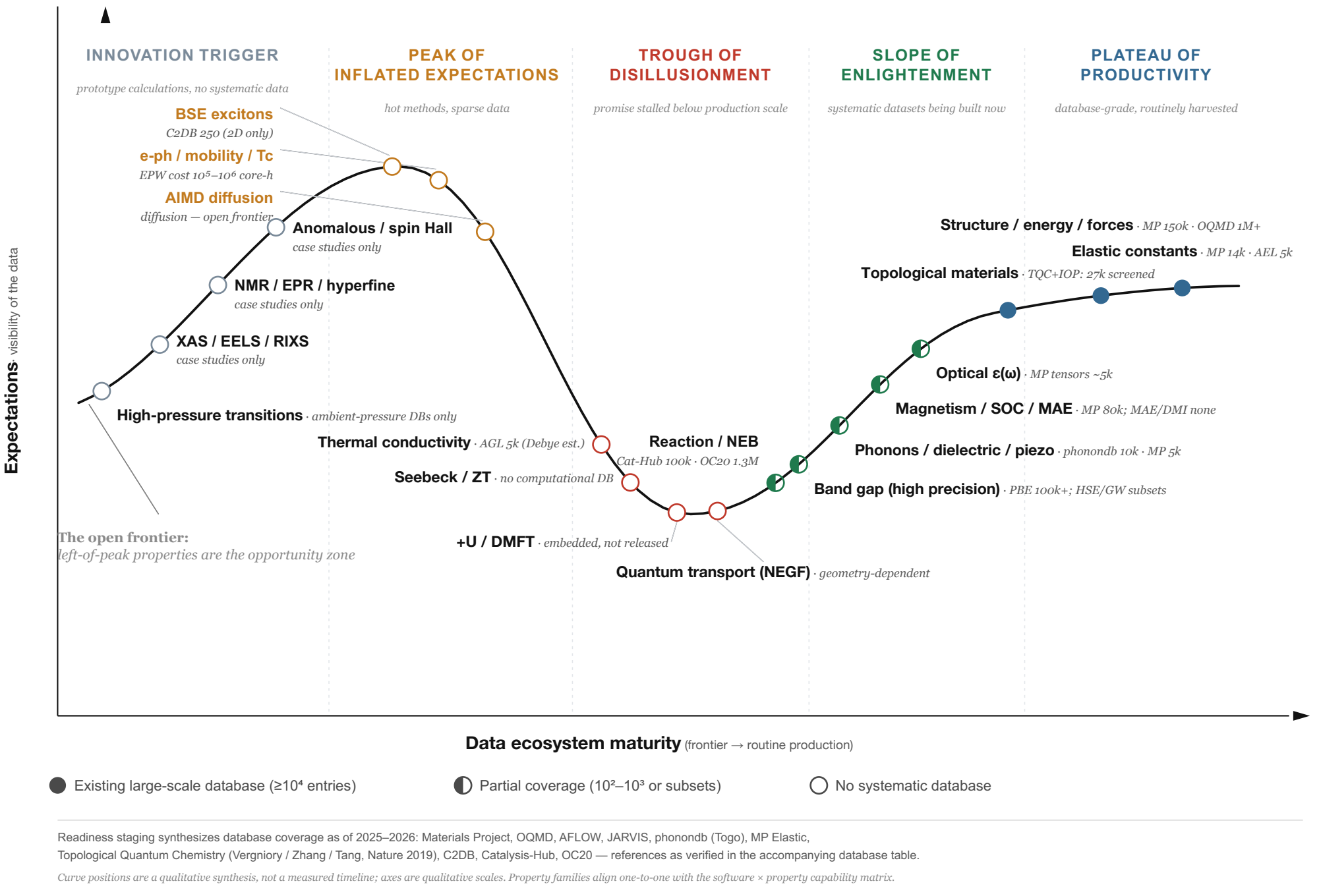

Figure 3 | The data readiness cycle of computational materials properties. Nineteen

property families are positioned by the maturity of their data ecosystems; marker fill again encodes database coverage. Properties left of the peak constitute the open frontier.

Three families have reached the plateau of productivity. Ground-state structure, energies, and forces are covered at the $10^5$–$10^6$ scale by MP (~200,000 compounds), OQMD (over one million), AFLOW (over three million), JARVIS-DFT (~80,000), and Atomly (over 340,000 compounds, with a band structure for nearly every entry) [1–5]. Elastic constants are available for roughly 14,000 materials in MP [29] and thousands more in AFLOW-AEL [30].

Most remarkably, topological classification — a property that barely existed a decade ago — was industrialized in a single sweep when three groups screened the entire known inorganic crystallographic inventory of ~27,000 materials using symmetry-based diagnostics, flagging thousands of previously unrecognized topological insulators and semimetals [31–33].

Topology demonstrates that when a property admits a cheap, automatable, high-credibility signature, the jump from zero coverage to near-complete coverage can happen in a single publication cycle.

On the slope of enlightenment, coverage is being systematically built right now. PBE-level band structures exist for $10^5$ materials [1,3], with hybrid and GW subsets emerging for curated collections — notably the C2DB database of two-dimensional materials, which provides GW quasiparticle gaps and Bethe–Salpeter optical spectra for a few hundred systems [34,35].

Phonon dispersions are covered for $10^3$–$10^4$ materials through the MP phonon effort [36] and the phonondb repository (~10,000 compounds) built with the phonopy infrastructure [37]; dielectric tensors for ~5,000 materials [38] and piezoelectric tensors for ~3,000 [39] follow the same DFPT pipeline.

Magnetic moments and orderings exist for ~80,000 MP entries [1], although anisotropies and Dzyaloshinskii–Moriya interactions remain unharvested. Reaction energetics has its own ecosystem: Catalysis-Hub hosts adsorption and reaction energies for well over 100,000 surface systems [40], and the OC20 dataset released ~1.3 million relaxation trajectories explicitly to train machine-learned potentials for catalysis [41].

The trough and the region left of the peak tell the opposite story. Thermal conductivity is represented only by quasiharmonic Debye-model estimates (AFLOW-AGL, ~5,000 entries [42]) — no database of Boltzmann-transport solutions from third-order force constants exists. Strongly correlated (+U/DMFT) parameters are computed routinely but buried inside workflows rather than published as data in their own right.

And nine families have essentially no systematic computational database at all: high-pressure phase transitions, X-ray absorption and electron energy-loss spectra, NMR/EPR and hyperfine parameters, anomalous and spin Hall conductivities, BSE excitons in bulk solids, electron–phonon properties including mobility and superconducting Tc, AIMD-scale diffusion, thermoelectric transport coefficients, and NEGF quantum transport.

The distribution is quantitatively stark: of the nineteen families, three enjoy large-scale coverage, six partial coverage, and nine are blank. Nearly half of the computable property space remains an open frontier.

It is worth pausing on what "blank" means in practice: not that the community cannot compute these properties — Fig. 1 and Table 1 show it can — but that no one has yet found the combination of cost, protocol robustness, and task definition that makes computing them *at scale* a rational investment.

## 5. Reading the frontier: why the blanks are blank

It is tempting to read the blank zones as a simple to-do list, but the reasons for the blanks are diagnostic of what a roadmap must solve. Four distinct bottlenecks recur.

**Cost.** EPW-grade electron–phonon calculations at $10^5$–$10^6$ core-hours per material cannot be multiplied across $10^5$ materials with any realistic computing budget; the same holds for third-order force constants and BSE. These families await surrogate models and algorithmic breakthroughs more than they await larger clusters.

**Credibility and protocol sensitivity.** NMR parameters, +U-dependent properties, and many-body corrections depend on methodological choices that resist one-size-fits-all automation. Publishing such data without machine-readable provenance of those choices risks polluting the literature rather than enriching it.

**Geometry and context dependence.** Quantum transport and catalytic reaction barriers are properties of devices and interfaces, not of bulk crystals; defining a meaningful "entry" is itself a research problem. The success of OC20 [41] shows the way: the community had to agree on a standardized task definition before a database could exist.

**Lack of an automatable signature.** Topology was conquered because symmetry indicators made classification cheap. The families still awaiting their "symmetry indicator moment" — spectral functions, hyperfine tensors, transport coefficients — need analogous proxies before high-throughput screening becomes sensible.

## 6. A roadmap for the next decade

We distill the analysis into a three-horizon roadmap, ordered by readiness rather than by importance.

**Horizon 1 (now to ~3 years): consolidate and connect.** Phonons, dielectric and piezoelectric tensors, and magnetic properties can be pushed from $10^3$–$10^4$ to $10^5$ scale with existing codes and workflows, at moderate cost and high credibility.

Equally important is interoperability: the same property is currently computed by different databases with different codes and settings, making cross-database learning hazardous; community-adopted metadata standards, in the spirit of the FAIR principles [43], would make existing data comparable across repositories without a single new calculation.

**Horizon 2 (~3–7 years): industrialize the mid-cost properties.** Hybrid/GW band gaps, full phonon dispersions, thermal conductivity, and defect properties should follow topology's playbook: find the cheap proxy, validate it against premium calculations on a curated subset, then screen at scale.

Machine-learned interatomic potentials are emerging as the decisive enabler here: they turn phonon dispersions and thermal conductivity — whose third-order force-constant calculations once cost $10^4$–$10^6$ core-hours per material — into high-throughput observables, promising a step change in data volume for exactly these families.

The scale-up is already visible: Li et al. [44] recently used a deep-learning interatomic potential to characterize over 640,000 crystalline structures and compute lattice thermal conductivities for more than 230,000 dynamically stable ones, producing in a single study a thermal-transport database orders of magnitude larger than everything previously available.

Machine-learned surrogates trained on existing premium subsets can already predict mid-cost properties with quantified uncertainty; the databases of the late 2020s should store predictions *and* uncertainties alongside premium reference values, clearly flagged.

**Horizon 3 (~7–15 years): conquer the high-cost frontier.** The blank nine families of Fig. 3 require genuine infrastructure innovation.

Universal machine-learned interatomic potentials, pre-trained on existing structure–energy–force data and fine-tuned on premium subsets, are already collapsing the cost of AIMD-scale dynamics, thermal conductivity, and defect sampling by orders of magnitude; autonomous computing infrastructure, and curated benchmark datasets linking calculation to measurement for the spectroscopy families, complete the picture.

The scientific literature itself will become part of the data infrastructure.

Domain-tuned language-model agents such as MatChat [45] have already demonstrated knowledge-base capabilities over the materials literature — answering domain questions with traceable references; the next step is scale.

Pipelines of this kind can extract structured, machine-readable data — compositions, measured properties, experimental conditions — from the published record in bulk, turning decades of papers into a queryable complement to the computed databases rather than an archive that humans must read one article at a time.

Governance — versioned releases, documented uncertainty, and credit for data publication — will decide whether the frontier becomes shared infrastructure or fragmented silos.

Two design principles cut across all three horizons. First, *publish the protocol with the number*: a property value without its methodological provenance is only half a datum. Second, *store uncertainty as a first-class quantity*: for the frontier families especially, an honest error bar is more valuable than an extra digit.

## 7. Conclusions

The field of inorganic computational materials databases has completed its first act: the methodological toolbox is mature (Fig. 1 and Table 1), the cheap-and-reliable properties have been harvested at scale, and topology has shown that entire property classes can be industrialized in a single cycle when the right signature exists (Figs. 2–3).

The second act will be defined by the nine blank property families — spectroscopies, electron–phonon physics, thermal and quantum transport, high-pressure and dynamical phenomena — where coverage is absent precisely because cost, credibility, and task definition remain unsolved.

The roadmap proposed here is deliberately conservative in its near term (consolidate, connect, standardize) and ambitious in its long term (surrogate-accelerated, autonomous, uncertainty-aware computation at scale). The community that built the first generation of databases property by property now has the opportunity — and the responsibility — to build the second generation as coherent, trustworthy, and complete infrastructure for materials science.

## Author Contributions

M.L., S.M. and W.W. drafted the manuscript. All authors contributed to the creation and preparation of this work, participated in the discussion and interpretation of the results, provided critical comments and suggestions during the writing process, and read and approved the final version of the manuscript. The authors declare no competing financial or personal interests that could have appeared to influence the work reported in this paper.

## Acknowledgements

This research is supported by National Key R&D Program of China (Grant No. 2024YFF0508500, 2021YFA1400200, and 2021YFA0718700), Chinese Academy of Sciences (Grant No. YSBR047).